%% file: main.tex
\documentclass[
    twocolumn,
    amsmath,amstex,amssymb,
    aps, pra,
    floatfix,
    longbibliography,
    reprint]{revtex4-2}

\usepackage{graphicx}
\usepackage{dcolumn}%
\usepackage{bm}
\usepackage{lipsum}
\usepackage{gensymb}
\usepackage{physics}
\usepackage{lipsum} 
\usepackage{braket}
\DeclareMathOperator{\sinc}{sinc}

\usepackage[markup=underlined]{changes}

\usepackage{hyperref}
\hypersetup{
  colorlinks   = true, 
  urlcolor     = blue, 
  linkcolor    = red, 
  citecolor   = red 
}

\usepackage[overload]{empheq}
\usepackage{diagbox}
\usepackage{tabularx,booktabs}
\newcolumntype{Y}{>{\centering\arraybackslash}X}

\begin{document}

\title{Tunneling Spectroscopy in Superconducting Circuit Lattices}

\author{Botao Du}
 \email{du245@purdue.edu}
\author{Qihao Guo}%
\author{Santiago L\'opez}
\author{Ruichao Ma}%
 \email{maruichao@purdue.edu}

\affiliation{%
Department of Physics and Astronomy, Purdue University, West Lafayette, IN 47907, USA }%

\date{May 5, 2025}

\begin{abstract}

We demonstrate tunneling spectroscopy of synthetic quantum matter in superconducting circuit lattices. We measure site-resolved excitation spectra by coupling the lattice to engineered driven-dissipative particle baths that serve as local tunneling probes. Using incoherent particle source and drain, we independently extract quasi-particle and quasi-hole spectra and reconstruct the spatial structure of collective excitations. We perform spectroscopy of a strongly interacting Bose-Hubbard lattice to measure changes in energy spectra for superfluid and Mott-insulator states at different densities and observe the effects of three-body interactions.
Our results provide a new toolset for characterizing many-body states in analog quantum simulators.

\end{abstract}

\maketitle

\section{Introduction}

Scanning tunneling spectroscopy has become an essential tool for investigating the electronic properties of quantum materials at the atomic scale \cite{Chen2007}. By varying the energy of the probe relative to the sample, tunneling spectroscopy facilitates the extraction of the local density of states for both electrons and holes, thereby providing critical insights into collective excitations in the material and their underlying interactions and correlations.
Such local spectroscopy measurements are pivotal for the investigation of superconductivity \cite{Fischer2007-dv, Wolf2011-xs}, quantum magnetism \cite{Qiu2021-ue}, and emergent topological states \cite{Yin2021-ma}.

With recent progress in quantum simulation, engineered quantum systems serve to emulate condensed matter models and study synthetic quantum matter with tunable control and precise readout. Ideas for realizing tunneling spectroscopy in such analog quantum simulators have been proposed for ultracold atoms \cite{Kollath2007-lf, Kantian2015-sy, Gruss2018-dc}, and injection spectroscopy has been recently demonstrated in a non-interacting synthetic lattice \cite{Paladugu2024-sx}.
In superconducting circuit quantum simulators, arrays of superconducting resonators and qubits host synthetic quantum matter comprised of interacting microwave photons \cite{Blais2004-nm, Houck2012-ol, Carusotto2020-ct}. Direct microwave spectroscopy measurements can be used to extract lattice parameters \cite{Ma2017-hw}, measure topological edge states \cite{Owens2018-gu}, or probe energy-resolved transport \cite{Fitzpatrick2017-sy, Fedorov2021-sr}. Using coherent local control, the many-body energy spectra can also be extracted from the time-evolution of initial product states \cite{Roushan2017-hh} or via many-body Ramsey experiments \cite{roberts2023manybody}. Nevertheless, it remains a challenge to perform site-resolved spectroscopy in strongly correlated lattices at the level of single collective excitations. Furthermore, the asymmetry between the quasi-particle and quasi-hole spectra provides essential insights into the underlying interactions of the constituent particles. However, most spectroscopic methods in synthetic quantum matter cannot distinguish particle excitations from hole excitations. 

In this work, we demonstrate tunneling spectroscopy in a superconducting circuit lattice by coupling the lattice to engineered driven-dissipative probes. The local probes serve as particle source and particle drain, as realized in our recent work \cite{du2024probing}, enabling particles to be injected or removed locally with controlled energy. We measure site-resolved quasi-particle and quasi-hole spectra of a strongly interacting Bose-Hubbard lattice, and reconstruct the spatial structure of quasi-particles from the site-resolved spectra. By performing the spectroscopy at different lattice fillings, we reveal changes in the many-body excitation spectra of different superfluid and Mott insulator states and directly observe the effects of multi-particle interaction. 
These experiments provide a new toolset for manipulating and characterizing quantum many-body states in superconducting circuits.

\begin{figure}[!t]
    \includegraphics[width=0.90\columnwidth]{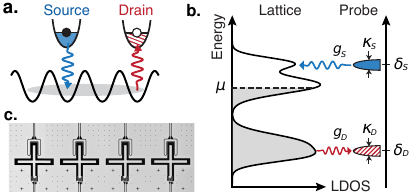}
    \caption{Illustration of site-resolved tunneling spectroscopy. (a) Locally coupled tunable particle baths act as tunneling probes. (b) The narrow energy bandwidth particle source (particle drain) measures the local density of states of unoccupied quasi-particle (quasi-hole) states. The many-body system is filled to an effective chemical potential $\mu$. (c) Image of the superconducting circuit lattice used in this work.
    }
    \label{fig:fig1} 
\end{figure}

\section{Local lattice excitation spectra}

Consider a tight-binding lattice model that hosts an interacting many-body state, illustrated in Fig.\,\ref{fig:fig1}. We perform tunneling spectroscopy by coupling lattice sites to particle baths with tunable narrow-band energy, and extract the local excitation spectra of the many-body state from the tunneling current between the lattice site and the probe as their energy detuning is varied. If the lattice is a closed system apart from its coupling to the probe, we can infer the tunneling current via the change in the total particle number in the lattice.
Here, we will employ two types of particle baths as the tunneling probe to access both the quasi-particle and quasi-hole spectra: If the probe is a particle source ($S$), particles are added to the many-body state at the energies of unoccupied quasi-particle states. If the probe is a particle drain ($D$), particles are removed from the many-body state at energies that match the unoccupied quasi-hole states.

For a many-body state $\psi$ with energy $E_\psi$, the local quasi-particle [quasi-hole] excitation spectrum at lattice site $i$ as a function of the probe frequency $\omega$ is:
\begin{equation}
I_i^\text{part[hole]}(\psi,\omega) = \sum_s \left| \mel{\phi_s}{a_i^\dagger[a_i]}{\psi} \right|^2 \delta(\frac{E_s-E_\psi}{\hbar} - \omega)
\label{eqn:1}
\end{equation}
$a_i^\dagger$($a_i$) is the particle creation (annihilation) operator on lattice site $i$, the sum in $s$ is over all many-body eigenstates of the lattice with eigenenergy $E_s$ and eigenfunction $\phi_s$, and $\hbar$ is the reduced Planck constant.
In essence, the excitation spectrum is the density of many-body states that can be reached from $\psi$ by adding or removing particles with frequency $\omega$ at site $i$. Here, we consider finite-size lattices with a discrete frequency spectrum.
In real systems, the frequency $\delta$-function will have finite widths due to intrinsic decoherence.

We now couple the lattice site $i$ to a local tunneling probe characterized by a narrow frequency bandwidth centered at $\omega$. We focus on the weak coupling regime, where the coupling rate between the lattice site and the probe is much smaller than the probe's bandwidth. We examine $N(t,\omega,i)$, the expectation value of the total number of particles in the lattice at time $t$, measured after switching on the probe coupling at $t=0$.
When the probe is a source/drain, particles enter/leave the lattice at energies near $\hbar \omega$ at a rate proportional to the local quasi-particle/quasi-hole spectrum integrated over relevant frequencies:
\begin{align}
\begin{split}    
    \dv{}{t} N(t, \omega, i) &= +\frac{4 g_\text{S}^2}{\kappa_\text{S}} \int d\omega' \, I_i^\text{part}(\psi(t),\omega') \,K(\omega-\omega',t) \\
    \dv{}{t} N(t, \omega, i) &= -\frac{4 g_\text{D}^2}{\kappa_\text{D}} \int d\omega' \, I_i^\text{hole}(\psi(t),\omega') \,K(\omega-\omega',t) 
\end{split} 
\label{eqn:2}
\end{align}
Here $g_\text{S/D}$ is the coupling rate between the lattice site and the particle source/drain, and $\kappa_\text{S/D}$ is the bandwidth of the source/drain. The many-body state $\psi(t)$ is written with explicit time dependence to indicate its evolution as a result of the probe. $K(\omega-\omega',t)$ is a frequency response function that describes the effective frequency resolution of the probe. The response function depends on the probe bandwidth, the probe-lattice coupling, and the finite probe duration. For a probe duration of $t$ with constant probe-lattice coupling, the Fourier-transform-limited frequency response is $K(\omega-\omega',t) = \sinc^2[(\omega-\omega') t/2]$ with a characteristic bandwidth of $2\pi/t$, which limits the frequency resolution of the tunneling spectroscopy at short times $t$. At longer times and in the weak coupling limit ($g_\text{S/D} \ll \kappa_\text{S/D}$), the frequency resolution is determined by the probes' frequency bandwidth $\kappa_\text{S/D}$. In our experiments, the probes are engineered from lossy microwave resonators, giving rise to Lorentzian frequency responses $K(\omega - \omega') = 1/[1+4(\omega - \omega')^2/\kappa_\text{D/S}^2]$.
Lastly, the proportional constants $4g_\text{S/D}^2/\kappa_\text{S/D}$ are the effective rates at which the source/drain adds/removes particles via a resonant quasi-particle/hole state of the lattice in the weak coupling limit. See related discussions in Ref.\,\cite{Kantian2015-sy} and additional details in Supplemental Material (SM) Sec.\,B \footnote{See Supplemental Material for device parameters, modeling, and additional analysis, which includes Refs.\,\cite{PhysRevLett.114.240501, Strand2013-wt, Pocklington2022-qr, groszkowski2021scqubits}.}.

\section{Spectroscopy of a circuit Bose-Hubbard lattice}

\begin{figure*}[!t]
    \includegraphics[width=1.7\columnwidth]{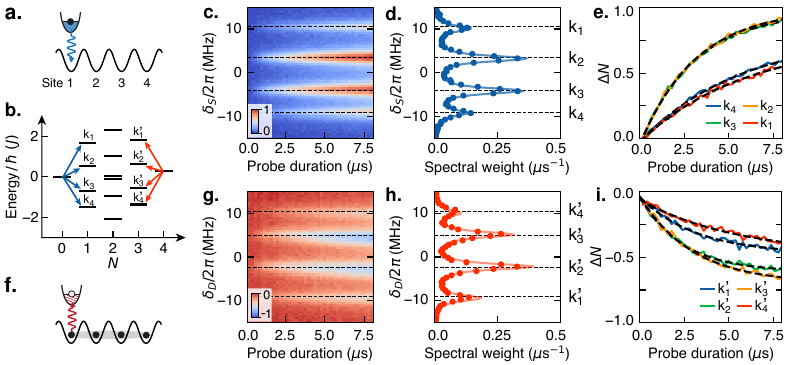}
    \caption{Measuring local excitation spectra of a Bose-Hubbard lattice. (a) Tunnel-coupled particle source probes quasi-particle excitations of the empty lattice, illustrated in the many-body spectrum (b). (c) Measured population change as a function of probe detuning and duration $\mathit{\Delta}N(\delta_\text{S},t)$, and (e) data and exponential fits at the four resonant detunings $\delta_\text{S}$. (d) The measured (dots) and numerically calculated (line) quasi-particle spectrum.
    (f) Tunnel-coupled particle drain probes quasi-hole excitations of the $n=1$ Mott state. (g) Measured probe-induced population change $\mathit{\Delta}N(\delta_\text{D},t)$, and (i) data and exponential fits at the four resonant $\delta_\text{D}$. (h) The measured and calculated quasi-hole spectrum. 
    Data in all figures are typically averaged over 10,000 experimental runs, with standard errors of the mean below 1\%. Other systematic uncertainties in the measured population are below $\pm$1\%, see \cite{du2024probing}.
    }
    \label{fig:fig2}
\end{figure*}

We demonstrate tunneling spectroscopy in a strongly interacting superconducting circuit Bose-Hubbard lattice comprised of a one-dimensional array of four transmon qubits where the particles are microwave photons (Fig.\,\ref{fig:fig1}c, see detailed device parameters and experimental setup in SM Sec.\,A and \cite{du2024probing}). The lattice is described by the Hamiltonian:
\begin{equation}
\begin{split}
 \mathcal{H}_{\text{BH}}/\hbar & = \sum_{<ij>}{J a_i^\dagger a_j} +\frac{U_2}{2}\sum_i{n_i(n_i-1)} \\
& +\frac{U_3}{6}\sum_i{n_i(n_i-1)(n_i-2)} + \sum_i \epsilon_i n_i 
\end{split}
\label{eqn:3}
\end{equation}
Here $a_i^\dagger$ is the bosonic creation operator for a microwave excitation on lattice site $i$, $J$ is the tunneling rate between lattice sites, $n_i=a_i^\dagger a_i$ is the on-site occupancy, $U_2$ ($U_3$) is the on-site two-body (three-body) interaction, and
$\epsilon_i$ is the on-site energy.
In our device, the effective on-site interactions are given by the energy level structure of the transmon qubit \cite{Koch2007-cz}, corresponding to strong attractive interactions of $U_2 \approx -2\pi\times 245$\,MHz and $U_3 \approx -2\pi\times 28$\,MHz. Experiments in this work are limited to on-site occupancy $n_i\leq 3$, so higher-order interactions beyond three-body are irrelevant. 
The fixed tunneling rate $J$ comes from capacitive coupling between the transmons, resulting in a uniform nearest neighbor tunneling $J_\text{NN} \approx  2\pi\times 5.9$\,MHz and a small residual next-nearest neighbor tunneling $J_\text{NNN} \approx  2\pi\times 0.5$\,MHz. In the remainder of this paper, we will use $J$ to denote the dominant $J_\text{NN}$ while $J_\text{NNN}$ contributes only small frequency shifts in the calculated lattice spectra.
The on-site energies $\epsilon_i$ are tunable within $2\pi\times (4.0-5.7)$\,GHz using local flux control lines. Microwave photons in the lattice have a typical relaxation rate of $\Gamma_1 = 1/T_1 \approx 1/(30\,\mu s) \approx 2\pi\times 5$\,kHz measured from single transmon lifetimes, and a local dephasing rate of $\Gamma_\phi = 1/T_2^* \approx 1/(2.5\,\mu s) \approx 2\pi\times 60$\,kHz. 

As demonstrated in our previous work \cite{du2024probing}, we create local tunable particle baths in a hardware-efficient way by inducing resonant coupling between the transmon lattice sites and their individual readout resonators. Parametric modulation of the transmon frequency at the sum (difference) of the transmon and resonator frequencies creates an effective incoherent particle source (drain), where the lossy readout resonator provides the source of dissipation. The strength and frequency of the modulation determine the tunable coupling strength $g_\text{D/S}$ and the frequency detuning between the probe and the lattice $\delta_\text{D/S}$. See SM Sec.\,B for details on the implementation of the particle baths and relevant parameters.
For experiments in this work, we choose $g_\text{D/S} \approx 2\pi\times0.25$\,MHz to be small compared to the resonator linewidth $\kappa_r \approx 2\pi\times 1.5$\,MHz. In this weak coupling limit, the incoherent particle baths represent tunneling probes that are effectively non-interacting and infinite; the baths have spectral width $\kappa_\text{D/S}$ equal to the resonator linewidth $\kappa_r$, which sets the frequency resolution for the tunneling spectroscopy. The spectral broadening from intrinsic lattice decoherence $\Gamma_1$ and $\Gamma_\phi$ are significantly less than $g_\text{D/S}$ and $\kappa_r$, so the $\delta$-function in Eq.\,\ref{eqn:1} remains valid.

With the present experimental parameters of $g_\text{D/S} \ll \kappa_\text{D/S} \ll J$ and a small size lattice, the change in the total particle number $\mathit{\Delta}N(t) = N(t)-N(t=0)$ as a function of the probe duration $t$ can be understood qualitatively  as follows: 
(i) For $t$ much shorter than the probe's dissipation timescale ($t \ll 2\pi/\kappa_r \approx 670$\,ns), the population change remains small and scales quadratically in time, $|\mathit{\Delta}N| \propto t^2 \ll 1$.
In this regime, the dominating dynamics is the coherent tunneling of particles between the lattice and the probe. Since the Fourier-limited spectral width $2\pi/t$ exceeds $\kappa_r$ during these short probe durations, the probe frequency can be near-resonant with several quasi-hole or quasi-particle excitations in the lattice. For each excitation resonant with the probe, the initial population transfer follows a Rabi oscillation given by $|\mathit{\Delta}N| = \sin^2(\Omega t)$ with Rabi rate $\Omega \approx g_\text{D/S}\,I_i(\omega)$. Since $g_\text{D/S} = 2\pi \times 0.25$\,MHz and $I_i(\omega) \leq 1$ characterizes the excitation’s transition matrix element, we find $\Omega t\ll 1$ for this short-time regime and therefore $|\mathit{\Delta}N| \approx \Omega^2 t^2 \ll 1$. When multiple resonances are present, $\mathit{\Delta}N$ remains approximately quadratic in time, representing the sum of several individual quadratic contributions.
(ii) For $t \geq 2\pi/\kappa_r \approx 670$\,ns, the probe's spectral resolution is no longer Fourier-limited, but is instead determined by $\kappa_r$. In our small lattice, where $\kappa_r\ll J$, the probe can resolve individual many-body transitions. Due to the strong on-site interaction $U_2$, the many-body excitation spectrum exhibits large nonlinearities that can exceed the probe's energy resolution. Consequently, if the probe is initially resonant with a particular quasi-particle (quasi-hole) excitation, the particle source (drain) can only add (remove) one particle before becoming off-resonant for subsequent excitations. In this regime, the population change in the lattice saturates towards $+1$ ($-1$) as $|\mathit{\Delta}N(t)| \approx 1-e^{-t/\tau}$, with a time constant given by $\tau \approx \kappa_r/(4g_\text{S(D)}^2 I_i(\omega))$. See SM Sec.\,C.3 for a detailed numerical example that illustrates these timescales.

\subsection{Quasi-particle and quasi-hole spectra}

We start by measuring the particle excitation spectrum of an initially empty lattice with all sites at the same frequency $\omega_\text{latt} \approx 2\pi\times 4.6$\,GHz (Fig.~\ref{fig:fig2}a). We couple the first lattice site to its local particle source with a coupling rate $g_\text{S}\approx 2\pi\times0.25$\,MHz and measure the total population after a variable probe duration of up to $t=8\,\mu$s. 
The individual readout resonators allow us to perform multiplexed readout of the on-site occupancy.
In Fig.~\ref{fig:fig2}c, we plot $\mathit{\Delta}N(t)$ measured at different probe detuning $\delta_\text{S}$.
We observe population growth at four distinct detunings, corresponding to the resonant population transfer from the particle bath into the four single-particle eigenstates of the lattice, labeled as $k_1$ to $k_4$. The single-particle states correspond to quasi-momentum eigenstates with energies $E(k_n)/\hbar = 2J \cos(k_n)$ and real-space wavefunctions $\psi_n(i) = \sqrt{\frac{2}{L+1}}\, \sin\left(k_n i\right)$, where the allowed quasi-momenta are $k_n = \frac{\pi n}{L+1}$ ($n = 1, 2, \dots, L$ with $L$ being the size of the lattice, and $i$ is the site index).
These single-particle excitation processes are illustrated in the many-body spectrum (Fig.~\ref{fig:fig1}b), here showing only states in the hard-core boson limit with on-site occupancy $n_i\leq 1$. On resonance with each transition, $\mathit{\Delta}N$ saturates towards 1. This saturation occurs because, with our experimental parameters, the rate at which the probe excites a second particle into the lattice is roughly two orders of magnitude lower than the single-particle excitation rates. This difference arises from the probe frequency mismatch within the non-linear spectrum and the limited wavefunction overlap.

To obtain the particle excitation spectra, we fit $\mathit{\Delta}N(t)$ at each detuning to an exponential function and extract the relative spectral weight from the slope of the fitted exponential at $t=0$. For the probe parameters here, the initial dynamics during which $\mathit{\Delta}N$ scales quadratically with time only contribute a small $\Delta N \ll 1$ so the exponential functions fit well to the measured data as seen in Fig.~\ref{fig:fig2}e. 
From the fits at the four resonant frequencies, $\mathit{\Delta}N$ saturates to values very close to 1 at $k_2$ and $k_3$. For $k_1$ and $k_4$, the fitted $\mathit{\Delta}N$ saturates towards $\approx 0.8$, due to the competition between the relatively weak pumping rate of the particle bath and the intrinsic relaxation in the lattice. 
In general, if the spectrum has (near-) degenerate multi-particle excitations, $\mathit{\Delta}N$ may not saturate towards 1 or follow a well-defined exponential at longer times. However, our analysis method can still yield an accurate spectrum if the exponential fit is applied only to data from early times when $\mathit{\Delta}N<1$.
The measured particle spectrum of the vacuum state is shown in Fig.~\ref{fig:fig2}d, in excellent agreement with the spectrum calculated from Eq.\,\ref{eqn:1}-\ref{eqn:2} without any free parameters. For the calculation, we use a Lorentzian with linewidth $\kappa_r$ and unity amplitude as the frequency response function $K$, and the eigenstates and eigenenergies are obtained from $\mathcal{H}_{\text{BH}}$ using the measured lattice parameters. 
While we measure and show the full dynamics of $\mathit{\Delta}N(t)$ here, only a few different probe durations are needed to extract the spectra accurately.

Next, we show the measurement of the quasi-hole spectrum. We choose the $n=1$ Mott insulator with one particle per lattice site as the initial state. In the Mott state, on-site particle number fluctuations are highly suppressed by the strong interaction $U_2$.
The $n=1$ Mott state is prepared by first applying individual resonant $\pi$-pulses to each transmon site while they are detuned from one another, followed by a rapid flux ramp that brings all sites into resonance at $\omega_\text{latt}$ within approximately 2\,ns. The ramp is fast enough to suppress population transfer via tunneling, yet slow enough to avoid non-adiabatic excitation to higher transmon levels.
We couple the first site to the local particle drain to perform the tunneling spectroscopy (Fig.~\ref{fig:fig2}f). Similar to the particle source above, the particle drain has an energy resolution given by the resonator linewidth $\kappa_r$ and we use a coupling strength of $g_\text{D} \approx 2\pi\times0.25$\,MHz.

For an initial state with finite population and probe duration up to $8\,\mu$s, the intrinsic relaxation in the lattice leads to population decay even when the probe is off-resonant from any transitions. We separately measure the intrinsic decay $N_0(t)$ with the probe off and subtract it from the measured $N(t)$ with the probe on to obtain the probe-induced population change $\mathit{\Delta}N(t) = N(t)-N_0(t)$, plotted in Fig.~\ref{fig:fig2}g as a function of the probe detuning $\delta_\text{S}$. 
In the hard-core limit, the Bose-Hubbard model exhibits particle-hole symmetry. The quasi-hole excitations of the $n=1$ Mott insulator are delocalized holes that resemble the delocalized single-particle excitations of the vacuum state with the same dispersion. The quasi-hole excitations appear at probe detunings that are reversed from the frequencies of the quasi-hole states in the many-body spectra, as seen in Fig.~\ref{fig:fig2}h. This is because a quasi-hole state with negative detuning is occupied by a particle with positive detuning during the tunneling spectroscopy. In our case, the single-particle/hole dispersions are slightly asymmetric as a result of the finite $J_\text{NNN}$ and the negative finite $U_2$. 
In Fig.~\ref{fig:fig2}i, we show the population change at each of the four quasi-hole resonances. The drain-induced population change saturates towards $\mathit{\Delta}N \sim -0.6$ for $k_2'$ and $k_3'$ and $\mathit{\Delta}N \sim -0.45$ for $k_1'$ and $k_4'$. These smaller $|\mathit{\Delta}N|$, compared to the expected value of 1, result from the intrinsic lattice relaxation. Without coupling to any probe, the transmon $T_1$ would lead to a population decay of $\mathit{\Delta}N_0 \approx -1$ in a duration of $8\,\mu$s starting from the Mott state. Since the probe is only resonantly coupled to one transition in the non-linear many-body spectrum, the drain-induced population change becomes suppressed as the initial Mott state decays.

In general, the intrinsic relaxation of many-body states is a complex cascaded state-dependent process. However, our measurement is robust to weak decoherence because we extract the spectral weights from the fitted slopes at early times when the population change due to intrinsic relaxation is small. 
We observe good agreement between the measured quasi-hole spectrum of the $n=1$ Mott state and the calculation (Fig.~\ref{fig:fig2}h). The measured spectrum has a slightly lower amplitude than the calculated spectra, with an overall scaling of $\approx 90$\%. We attribute this to state preparation errors that result in a reduced initial population in the $n=1$ Mott state.

\subsection{Imaging site-resolved wavefunctions}

\begin{figure}[!t]
    \includegraphics[width=0.98\columnwidth]{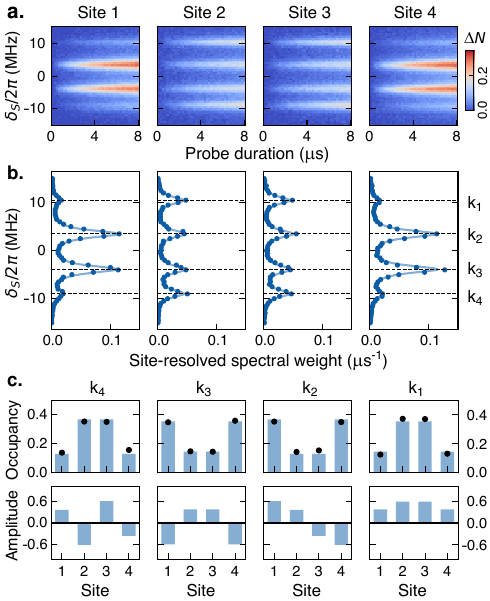}
    \caption{Reconstruct single-particle wavefunctions using site-resolved spectra. (a) Population change on each lattice site for the experiment in Fig.~\ref{fig:fig2}a-e. (b) Measured spectra from data in (a). Lines are fit to a sum of Lorentzians with width $\kappa_r$ to obtain the site-resolved spectral weights for each single particle mode $k_1$ to $k_4$. (c) \textit{Top:} On-site occupancy of the single-particle eigenstates from the normalized spectral weights. Dots are data, and bars are from calculated eigenstates. \textit{Bottom:} Calculated on-site amplitudes of the corresponding wavefunctions.
    }
    \label{fig:fig3}
\end{figure}

\begin{figure*}[!htb]
    \includegraphics[width=1.98\columnwidth]{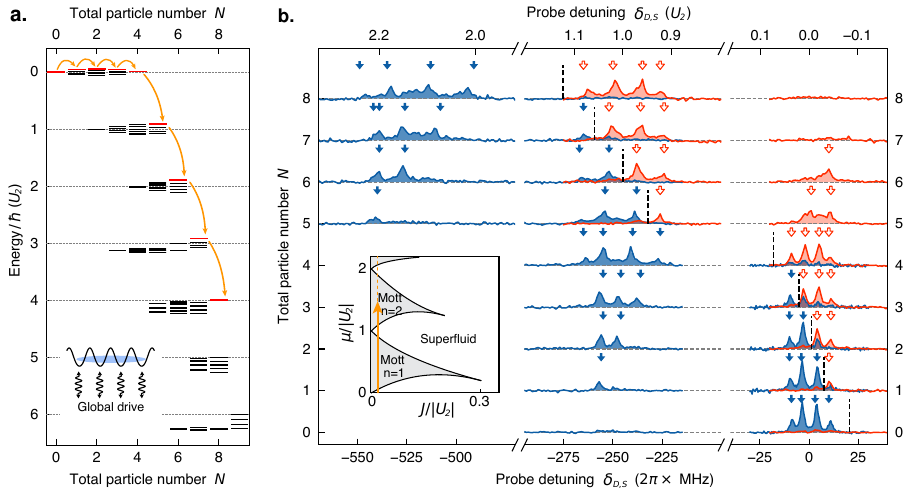}
    \caption{Density-dependent tunneling spectroscopy. (a) Many-body spectra of the Bose-Hubbard lattice beyond the hard-core limit. We prepare the highest-energy states for $N$ up to 8 (red thick lines) with a time-varying global coherent drive (inset) that sequentially populates the lattice via many-body Landau-Zener transitions (orange arrows). 
    (b) The measured quasi-particle spectra (blue with dark fill) and quasi-hole spectra (red with light fill) at each lattice population $N$, probed at site 1. The dotted vertical lines indicate the end detuning of the coherent drive used to prepare each state, which acts as the effective chemical potential $\mu$. Arrows above the spectra show frequencies of calculated excitations. The inset in (b) illustrates the phase diagram, and the vertical arrow indicates the states we accessed. The regions behind the insets contain no spectral features.
    }
    \label{fig:fig4} 
\end{figure*}

In addition to the spectral information, the site-resolved readout in our circuit Bose-Hubbard lattice also provides spatial information of the quasi-particle and quasi-hole excitations. As a simple demonstration, we measure the on-site occupancy of single-particle eigenstates using the same tunneling spectroscopy experiment described in Fig.~\ref{fig:fig2}a-e. In Fig.~\ref{fig:fig3}a, we plot the particle number change on each lattice site during the tunneling spectroscopy. We use the same exponential fitting methods discussed previously to obtain the site-resolved quasi-particle spectra shown in Fig.~\ref{fig:fig3}b. We fit the spectra with sums of Lorentzian functions with fixed linewidth $\kappa_r$; the amplitudes of the Lorentzian give the spectral weight at each site $i$ and for each eigenstate $k_j$. For each of the four single-particle eigenstates, the site-resolved spectral weights are directly proportional to the on-site occupancies of the wavefunction. In Fig.~\ref{fig:fig3}c, we show the normalized on-site occupancies for the four single-particle quasi-momentum eigenstates, in agreement with the calculated densities from the single-particle wavefunctions.

Here, we used the site-resolved readout with a spatially fixed local probe to obtain the spatial density distributions of collective excitations in the lattice. Alternatively, we can perform tunneling spectroscopy with the probe coupled to different lattice sites and measure the spatial distribution from the local spectral weights. Our hardware-efficient implementation of the tunneling probe is well suited for such analogous scanning tunneling spectroscopy.
Beyond simple single-particle states, the tunneling probe needs sufficient energy resolution to excite and measure a single collective excitation in a generic many-body state. Hence the method demonstrated here can be applied to gapped phases or finite-size systems with large nonlinearities to reveal spatial information of collective excitations.

\subsection{Spectroscopy across the superfluid to Mott insulator transition}

We now perform the tunneling spectroscopy at different filling densities of the Bose-Hubbard lattice. As the many-body state undergoes density-induced transitions between strongly correlated superfluids and Mott insulators, we use the measured excitation spectra to identify changes in the energy gap in different quantum phases and observe the effects of multi-particle interactions.

We use a global coherent microwave drive to adiabatically populate the lattice via many-body Landau Zener transitions \cite{du2024probing}, illustrated in Fig.\,\ref{fig:fig4}a. A similar method was used to coherently prepare entangled states in a two-dimensional circuit Bose-Hubbard lattice \cite{Karamlou2024-gt}. Starting with an empty lattice, we turn on the global drive at frequency $\omega_\text{drive}$ to a Rabi rate of $\Omega_\text{drive} = 2\pi\times 4.0$\,MHz in 300\,ns with an initial detuning $\omega_\text{drive}-\omega_\text{latt} = 2\pi\times 30\,\text{MHz}\approx 5J$. We then sweep $\omega_\text{drive}$ at a rate of $-2\pi\times 80\,\text{MHz}/\mu$s to different end drive frequencies $\omega_\text{drive}^\text{end}$, before turning off the drive in 300\,ns. Using drive sweeps with end detunings $\omega_\text{drive}^\text{end}-\omega_\text{latt} = 2\pi\times \{7.6,1.2,-5.2, -18\}\,\text{MHz}\approx \{1.2J, 0.2J, -0.9J, -3J\}$, we prepare the highest energy eigenstates with total particle number $N=\{1,2,3,4\}$ respectively. The $N=4$ state is the $n=1$ Mott insulator. Here, the end drive detuning acts as an effective chemical potential $\mu$ that determines the total particle number in the lattice. 
Starting from the state with $N=4$, we apply a second global coherent drive to prepare states with $N>4$ beyond the hard-core boson limit. We start at drive detuning $-2\pi\times 205\,\text{MHz} \approx U_2 +6.7J$ and stop at end detunings of $2\pi\times \{-232.2,-245,-259.4,-275.4\}\,\text{MHz} \approx \{U_2 + 2J, U_2, U_2 -2.4J, U_2 - 5J\}$, to populate the highest energy eigenstates with $N=\{5,6,7,8\}$ respectively. The $N=8$ state is the doubly-filled $n=2$ Mott insulator. States other than $N=4$ and $8$ are strongly correlated superfluids that exhibit large on-site number fluctuations and interaction-induced density-density correlations \cite{Saxberg2022-tt}. 
Our transmon lattice has $U_2<0$, representing a strongly attractive Bose-Hubbard model. By preparing the highest energy states in the finite-size lattice, we effectively study the physics of the ground states of the strongly repulsive Bose-Hubbard model.
As $N$ increases from 0 to 8, we traverse a path in the phase diagram of the 1D Bose-Hubbard model illustrated in the inset of Fig.~\ref{fig:fig4}b (see for example the predicted phase diagram in Ref.~\cite{Ejima2011-df}). 

The measured density-dependent quasi-particle and quasi-hole spectra are shown in Fig.~\ref{fig:fig4}b. For $N=0$, the particle spectrum has been shown previously in Fig.~\ref{fig:fig2}a. The hole spectrum shows small finite amplitudes at the same single-particle eigenstate frequencies; these result from the equilibrium thermal population in the lattice which is measured to be approximately 5\% per site. 
For $N=1,2,3$, the superfluid states show low-frequency quasi-hole and quasi-particle excitations in the ground band with probe detunings $|\delta_\text{D,S}| \lesssim 2J$. The observed asymmetry between the particle and hole spectra is a signature of the strong interactions. In addition, states with $N\geq 1$ can be excited to states in the second band with two particles on the same site, which appear in the quasi-particle spectra near $\delta_\text{S} \sim U_2$ due to the on-site two-body interaction. 

At $N=4$ (the $n=1$ Mott state), all quasi-hole excitations appear in the ground band while the quasi-particle excitations are separated by $\sim U_2$ in the second band. This energy separation is the characteristic Mott gap of the insulating Mott phase. The quasi-particle states in the second band contain one delocalized particle tunneling on top of a unity-filled $n=1$ Mott state, which tunnels with a bosonic enhanced rate of $2J$. As a result, the excitations in the second band have a larger frequency dispersion spanning $|\delta_\text{D,S} - U_2| \lesssim 2\times 2J$. The small residual amplitude of the particle excitation peaks in the ground band is from imperfect preparation of the $n=1$ Mott state due to thermal population and decoherence during the adiabatic drive.

The states we prepared at $N=5,6,7$ are once again correlated superfluids, while $N=8$ is the $n=2$ Mott insulator.
Beyond the ground and second band excitations, these states with $N\geq 5$ can be excited to states in the third band with three particles on the same site. A third particle on a lattice site will experience both the two-body interaction $U_2$ and three-body interaction $U_3$, so these quasi-particle resonances appear near probe detunings $\delta_\text{S} \sim 2U_2+U_3$. For example, the quasi-particle excitations of the $N=8$ Mott state correspond to a delocalized particle with a bosonic-enhanced tunneling rate of $3J$ on top of the doubly-filled Mott state. As seen in the plotted spectra, these excitations to the third band are observed at probe frequencies in the range of $|\delta_\text{S} - 2U_2 - U_3| \lesssim 3\times 2J$.
Therefore, our measurements provide a direct spectroscopic probe of the dispersion and interactions of the transmon lattice beyond the hard-core limit. Utilizing higher levels of transmons opens new possibilities for analog quantum simulation \cite{PRXQuantum.3.040314}. For instance, the realization of lattice models with dominant $U_3$ interaction using effective cancellation of $U_2$ \cite{Cardarelli2016-zf} could lead to strongly-correlated phases with charge-density-wave order \cite{Daley2014-tr}.

\section{Discussion and Outlook}

The tunneling spectroscopy demonstrated in this paper offers several advantages over existing spectroscopic methods in superconducting circuit lattices. Conventional spectroscopy relies on weak and long coherent drives to achieve high energy resolution, where the absorption/emission signals result from the balance between the external drive and intrinsic dissipation \cite{Blais2021-bn}. The exact rates of the driving or dissipation are difficult to characterize a priori and generally vary for different many-body states, thereby complicating the extraction of accurate spectral weights. In our method, the tunneling rate is set by the controlled coupling to the incoherent probe, making the obtained spectra insensitive to weak intrinsic decoherence. Moreover, the ability to distinguish between quasi-hole and quasi-particle excitations, as we demonstrated, remains challenging in conventional coherent spectroscopy.

Our locally coupled probes enable site-resolved tunneling spectroscopy that is well-suited for investigating systems with spatially varying excitations. Examples include studying many-body localized systems by measuring the local level statistics and the spatial extent of rare Griffiths regions \cite{Abanin2019-lg, Agarwal2017-qa}; understanding strongly correlated topological models by probing edge states in hard-core Su–Schrieffer–Heeger models \cite{de-Leseleuc2019-yc} or interacting chiral edge states in a Chern insulator \cite{Owens2022-ua}; and characterizing soliton states in interacting lattices \cite{Blain2023-qw}.

In this paper, we used the multiplexed readout to measure the population across the entire lattice to infer the probe-induced tunneling current. To further reduce the measurement overhead, we can directly measure the tunneling current between the probe and the lattice \cite{du2024probing}.
Instead of the incoherent particle baths in this work, coherent two-level systems coupled to the lattice could also serve as tunneling probes \cite{Stenger2022-pt, Kantian2015-sy}.

Another potential extension is the use of multiple local probes: Employing both a source and a drain would enable energy- and site-resolved pump-probe spectroscopy to provide insights into the non-equilibrium dynamics. The dynamics of collective excitations can be accessed by varying the delay between the bath-induced excitation and a subsequent spectroscopy measurement.
Momentum-resolved spectroscopy has also been proposed, where arrays of local probes are coupled to the lattice to extract the spectral function \cite{Zawadzki2019-hz}.

Finally, our experiments open the door for the dissipative preparation of quantum many-body states using incoherent particle baths \cite{Ma2019-Mott}. The spectrally narrow baths can be used to populate strongly correlated states sequentially \cite{Umucalilar2017-ak}, and combining the energy-selective particle source and drain enables the autonomous stabilization of many-body phases in photonic synthetic quantum matter \cite{Lebreuilly2018-bs}.

\section*{Acknowledgments}

This work was supported by grants from the National Science Foundation (award number DMR-2145323) and the Air Force Office of Scientific Research (award number FA9550-23-1-0491). Part of this material is based upon work supported by the U.S. Department of Energy, Office of Science, National Quantum Information Science Research Centers, Quantum Science Center.


%

\input{supp_material.tex}

\end{document}

%% file: supp_material.tex

\renewcommand{\thetable}{S\arabic{table}}
\renewcommand{\theequation}{S\arabic{equation}}
\renewcommand{\thefigure}{S\arabic{figure}}
\setcounter{equation}{0}
\setcounter{figure}{0}
\setcounter{secnumdepth}{3}

\newpage

\onecolumngrid

\renewcommand{\appendix}{\par
  \setcounter{section}{0}
  \setcounter{subsection}{0}
  \setcounter{subsubsection}{0}
  \gdef\thesection{\Alph{section}}
  \gdef\thesubsection{\Alph{section}.\arabic{subsection}}
  \gdef\thesubsubsection{\Alph{section}.\arabic{subsection}.\roman{subsubsection}}
}

\appendix

\clearpage

\begin{center} 
    \large {\textbf {Tunneling Spectroscopy in Superconducting Circuit Lattices\\}}
    \vspace{0.1in}
    \uppercase{\textbf {Supplemental Material}}
    \vspace{0.1in}
\end{center}


\twocolumngrid

\section{Device parameters}

Experiments in this work are performed on the same device used in our previous work \cite{du2024probing}. Table \ref{tab:SI_qubit-param} lists key device parameters relevant to the current work.

The coherence times and lattice parameters are measured at the lattice frequency $\epsilon_i = \omega_{q}^\text{latt}$. $T_1$ and $T_2^*$ are measured from single-qubit relaxation and Ramsey experiments, nearest-neighbor tunnelings $J_{i,i+1}$ from coherent double-well oscillations between neighboring pairs of qubits, and on-site interactions $U_2$ and $U_3$ from single-qubit spectroscopy (see Fig.\,\ref{fig:figSI-gefh} and discussion below). The next-nearest-neighbor tunnelings $J_{i,i+2}$ are extracted from measured eigenenergies of the degenerate lattice; the values match estimates from finite element simulations of the device. When all qubits are tuned to the lattice frequency, we estimate residual lattice disorder to be $\lesssim 2\pi\times 0.1\,\text{MHz}$.
The qubit tuning range $\omega_{q_i}^{\textrm{min}}, \omega_{q_i}^{\textrm{max}}$, qubit to readout resonator coupling $g$ and resonator linewidth $\kappa_r$ are extracted from the measured flux-dependent qubit and resonator spectra. 

All qubits have finite thermal population $n^q_\text{th} \approx 5\%$ measured from ef-Rabi oscillations \cite{PhysRevLett.114.240501}, i.e. the equilibrium thermal ground state is $\rho_g^\text{th} \approx (1-n^q_\text{th})\ketbra{g}{g} + n^q_\text{th}\ketbra{e}{e}$. Thermal population in $\ket{f}$ ($n=2$) is negligible. The qubit population in the experiments is measured by averaging typically 10,000 runs, calibrated to readout signals from the qubit basis states $\rho_g^\text{th}$ and $\rho_e^\text{th}$. The latter is prepared by applying a resonant $\pi$ pulse to $\rho_g^\text{th}$. Since both basis states contain thermal population, we scale the measured average population by $1/(1-2\,n^q_\text{th})$ to arrive at the actual qubit population, e.g. $P_{\ket{g}}$ or $P_{\ket{e}}$, which is the quantity presented for all data in the main text.

We also measure the thermal population in the readout resonators to be approximately $2\%$, which has a negligible effect on the present experiments. 

\begin{table}[ht]
	{\renewcommand{\arraystretch}{1.4}
	\begin{tabularx}{0.4 \textwidth}{| Y  Y | Y | Y | Y | Y |} 
    
        \hline ~ & ~ & $Q_1$ & $Q_2$ & $Q_3$ & $Q_4$   \\ \hline\hline
        \multicolumn{2}{|c|}{$n^q_\text{th}$ } & 0.05 & 0.05 & 0.05 & 0.05 \\ \hline
        \multicolumn{2}{|c|}{$T_1$ ($\mu$s)} & 21 & 30 & 30 & 27  \\ \hline
        \multicolumn{2}{|c|}{$T_2^*$ ($\mu$s)} & 1.9 & 2.7 & 1.8 & 1.4 \\ \hline
        
        \multicolumn{2}{|c|}{$\omega_q^{\textrm{latt}}/2\pi$ (GHz)} & 4.6 & 4.6 & 4.6 & 4.6  \\ \hline

        \multicolumn{2}{|c|}{$U_2/2\pi$ (MHz)} & -245.6 & -245.6 & -246.6 & -245.7  \\ \hline

        \multicolumn{2}{|c|}{$U_3/2\pi$ (MHz)} & -28.6 & -28.2 & -25.7 & -28.3  \\ \hline
        
        \multicolumn{2}{|c|}{$J_{i,i+1}/2\pi$ (MHz)} & 5.9 & 5.9 & 5.9 & n/a  \\ \hline

        \multicolumn{2}{|c|}{$J_{i,i+2}/2\pi$ (MHz)} & 0.05 & 0.05 & n/a & n/a  \\ \hline  

        \multicolumn{2}{|c|}{$\omega_q^{\textrm{min}}/2\pi$ (GHz)} & 3.90 & 3.90 & 3.90 & 3.90  \\ \hline
        \multicolumn{2}{|c|}{$\omega_q^{\textrm{max}}/2\pi$ (GHz)} & 5.78 & 5.78 & 5.73 & 5.73  \\ \hline
        \multicolumn{2}{|c|}{$\omega_r^{\textrm{bare}}/2\pi$ (GHz)} & 6.320 & 6.273 & 6.228 & 6.176  \\ \hline

        \multicolumn{2}{|c|}{$g/2\pi$ (MHz)} & 67 & 65 & 64 & 65  \\ \hline
        
        \multicolumn{2}{|c|}{$\kappa_{\textrm{r}}/2\pi$ (MHz)} & 1.5 & 1.5 & 1.5 &  1.5  \\ \hline
    \end{tabularx}
    }
\caption{Device parameters.}
\label{tab:SI_qubit-param}
\end{table}

\subsection{Onsite interactions \label{sec:U3}}

A transmon qubit can be viewed as an anharmonic oscillator where the anharmonicity of the higher levels corresponds to effective on-site interactions between microwave photons in the qubit. Consider the lowest 4 Fock states $\vert 0 \rangle, \vert 1 \rangle, \vert 2 \rangle$ and $\vert 3\rangle$ (i.e. transmon $\ket{g}, \ket{e}, \ket{f}$ and $\ket{h}$ states), the qubit transition frequencies can be written using the effective two-body and three-body interactions $U_2$ and $U_3$ as: $\omega_{12} = \omega_{01} + U_2$ and $\omega_{23} = \omega_{01} + 2U_2 + U_3 = \omega_{12} + U_2 +U_3$.

\begin{figure}[ht]
    \includegraphics[width=0.85\columnwidth]{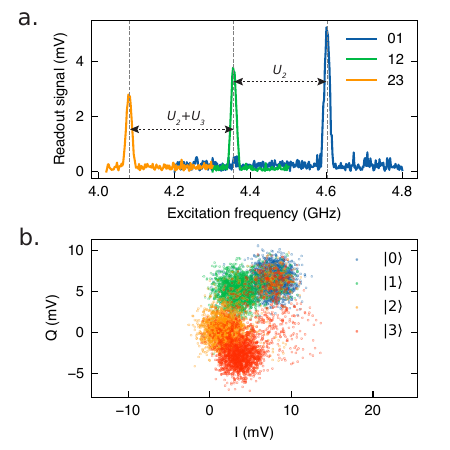}
    \caption{Characterization of transmon higher levels. (a) Single qubit spectroscopy for extracting the on-site $U_2$ and $U_3$ by measuring the transmon transition frequencies of Q1. (b) Readout signal for Q1 prepared in states $\vert0\rangle, \vert1\rangle, \vert2\rangle, \vert3\rangle$.
    }
    \label{fig:figSI-gefh}
\end{figure}

We extract these interaction strengths for each qubit individually at the lattice frequency, by measuring the transmon transition frequencies using both qubit spectroscopy and Ramsey experiments. For example, we show the coherent two-tone qubit spectroscopy data from qubit Q1 in Fig.\,\ref{fig:figSI-gefh}(a). In Fig.\,\ref{fig:figSI-gefh}(b), we show the typical qubit readout signal.

We also compare the measured transmon transition frequencies with theoretical predictions. We include the full transmon Hamiltonian and the coupled readout resonator in our numerical model:
\begin{align*}
    \mathcal{H}_{\text{transmon}} &= 4E_\text{C}(\hat n - n_g)^2 - \frac{1}{2} E_\text{J}\sum_n (\vert n \rangle \langle n+1\vert + \text{h.c.}) \\
    \mathcal{H} &=  \mathcal{H}_{\text{transmon}} + \hbar\omega_r b^\dagger b + \hbar g \,\hat n(b+b^\dagger)
\end{align*}
We numerically diagonalize $\mathcal{H}$ using \texttt{scQubits} \cite{groszkowski2021scqubits} and have verified that all measured parameters and transition frequencies in Table\,\ref{tab:SI_qubit-param} are in good agreement with theory. For example, the extracted transmon parameters for Q1 are $E_\text{J}/h = 13.38\, \text{GHz}$ and $E_\text{C}/h = 0.218 \,\text{GHz}$.

\section{Engineering local tunable particle baths}

\subsection{Qubit-resonator parametric coupling}

We consider the effective coupling between a flux-tunable transmon and its readout resonator when the transmon frequency is modulated by an AC flux signal. The qubit and resonator are capacitively coupled with static coupling $g$ and detuning $\Delta = \omega_r-\omega_q \gg g$.
We consider qubit frequency modulation $\omega_q(t) = \omega_q^0 + A_\text{mod} \cos{(\omega_\text{mod} t)}$. For $\omega_\text{mod} \approx \omega_r-\omega_q^0$ (red sideband), the modulation induces resonant coupling between the qubit and the lossy readout resonator---the qubit excitation tunnels to the resonator and decays via $\kappa_r$, so the red sideband modulation realizes a local narrow-band particle drain ($D$) for the lattice. When $\omega_\text{mod} \approx \omega_r+\omega_q^0$ (blue sideband), the modulation corresponds to a coherent two-photon drive that adds and removes pairs of excitations---the photon added to the resonator is lost via $\kappa_r$ leaving the other photon added to the qubit, turning the blue sideband modulation into a narrow-band incoherent source ($S$) of particles for the lattice. These tunable particle baths were implemented in our previous work to induce and measure quantum transport in a circuit lattice \cite{du2024probing}.

Following similar derivations in Refs.\,\cite{Strand2013-wt, Pocklington2022-qr}, we start with the Hamiltonian of the qubit-resonator system in the lab frame,
\[
\mathcal{H}/\hbar = \omega_q(t)a^\dagger a + \omega_r b^\dagger b + g(a^\dagger + a)(b^\dagger + b) + \eta[a^\dagger a]
\]
Here $b^\dagger$ is the creation operator of the resonator. $\eta$ is a polynomial function of local density $n = a^\dagger a$ that represents the nonlinearity of the transmon higher-levels, or equivalently, the onsite interactions in the Bose-Hubbard description.

\textbf{Particle drain:} For red sideband modulation with $\omega_\text{mod} \approx \omega_r-\omega_q^0$, the dynamics is best described in the rotating frame by applying a unitary transformation $\mathcal{H'} = \mathcal{\hat{U}^\dagger} \mathcal{H} \mathcal{\hat{U}} - i\mathcal{\hat{U}^\dagger} \frac{d}{dt}\mathcal{\hat{U}}$, where
\[
\mathcal{\hat{U}} = \exp\{ i (\omega_\text{mod} + \omega_q^0) t\, b^\dagger b \ + i (\omega_q^0 t + \frac{A_\text{mod}}{\omega_\text{mod}}\sin{(\omega_\text{mod}t)}) a^\dagger a \}
\]
Applying the rotating wave approximation, the effective Hamiltonian for the particle drain ($D$) reads:
\[ \mathcal{H}_\text{D} = \mathcal{H'} \approx \delta_\text{D} b^\dagger b + g_\text{D}(a^\dagger b + a b^\dagger) + \eta(a^\dagger a) \]
The effective resonant coupling between the particle drain and the qubit is $g_\text{D} = g\times J_1 \left( \frac{A_\text{mod}}{\omega_\text{mod}}\right)$, where $J_1$ is the 1st order Bessel function of the first kind. The effective detuning between the particle drain and the qubit lattice is $\delta_\text{D} = (\omega_r - \omega_\text{mod}) - \omega_q^0$.

For all experiments in this work, we operate in the weak-coupling limit of $g_\text{D} \ll \kappa_r$ where the
Born-Markov approximation is valid. In this limit, the resonator remains close to its ground state and can be adiabatically eliminated to arrive at the effective dynamics for the qubit. For simplicity, consider the resonant case $\delta_\text{D}=0$. The Heisenberg equations of motion in the rotating frame reads:
\[ \dot a = -\,i\,g_\text{D}\,b,\quad
\dot b = -\,i\,g_\text{D}\,a - \tfrac{\kappa_r}{2}\,b + (\mathrm{noise})\]
We set $\dot{b}\approx 0$, and consider only expectation values so the noise term that includes quantum fluctuations in the resonator averages to zero. It follows that $\expval{b}\approx -\tfrac{2\,i\,g_\text{D}}{\kappa_r}\,\expval{a}$, and substituting into the equation for $\dot a$ gives
$\expval{\dot a} \approx -\,\tfrac{2\,g_\text{D}^2}{\kappa_r}\,\expval{a}$.
Hence the qubit amplitude $\expval{a}$ decays exponentially with a time constant of $\kappa_r/2g_\text{D}^2$ while the qubit population $\expval{n}$ decays exponentially with a time constant of $\kappa_r/4g_\text{D}^2$. In this limit, the energy bandwidth of the particle drain is given by the linewidth of the resonator $\kappa_r$. 

\textbf{Particle source:} For blue sideband modulation with $\omega_\text{mod} \approx \omega_r+\omega_q^0$, the effective Hamiltonian is derived by going to a rotating frame using
\[
\mathcal{\hat{U}} = \exp\{ i (\omega_\text{mod} - \omega_q^0) t\, b^\dagger b \ + i (\omega_q^0 t + \frac{A_\text{mod}}{\omega_\text{mod}}\sin{(\omega_\text{mod}t)}) a^\dagger a \}
\]
and applying the rotating wave approximation. The effective Hamiltonian for the particle source ($S$) reads:
\[ \mathcal{H}_\text{S} = \mathcal{H'} \approx -\delta_\text{S} b^\dagger b + g_\text{S}(a^\dagger b^\dagger + a b) + \eta[a^\dagger a] \]
The effective coupling $g_\text{S} = g\times J_1\left( \frac{A_\text{mod}}{\omega_\text{mod}}\right)$, and effective detuning between the particle source and the qubit lattice is $\delta_\text{S} = (\omega_\text{mod} - \omega_r) - \omega_q^0$. Similarly, in the weak coupling limit where $g_\text{S} \ll \kappa_r$, the resonator can be adiabatically eliminated. The qubit population saturates towards 1 (excited state) with a time constant given by $\kappa_r/4\,g_\text{S}^2$.

\textbf{In summary:} The red/blue sideband flux modulation realizes local tunable particle drain/source with effective coupling $g_{\text{D/S}}$, linewidth $\kappa_r$, and effective detuning $\delta_\text{D/S}$. In the weak coupling limit ($g_\text{D/S} \ll \kappa_r)$, the particle drain/source acts like Markovian bath that removes/adds particles at a rate of $4g_\text{D/S}^2/\kappa_r$.

\subsection{Experimental tuning range of $g_\text{D}$ and $g_\text{S}$}

The effective parametric coupling strength $g_\text{D/S}$ is a function of the static qubit-resonator coupling, and the modulation amplitude and modulation frequency. With our device parameter shown in Table \ref{tab:SI_qubit-param}, the modulation amplitude is limited by the transmon frequency tuning range of $\sim 2\pi\times (1.8-1.9)$\,GHz and the qubit-resonator static coupling $g \sim 2\pi\times 65$\,MHz. For the particle source: applying the blue sideband flux modulation at $\omega_\text{mod} \approx 2\pi \times 10.8$\,GHz on qubit Q4, we achieve a maximum coupling rate of $g_\text{S} \approx 2\pi \times 2.4$\,MHz. For the particle drain, applying the red sideband flux modulation with $\omega_\text{mod} \approx 2\pi\times 1.7$\,GHz qubit on Q1, we get a maximum coupling rate of $g_\text{D} \approx 2\pi \times 12$\,MHz.

\section{Numerical modeling of data}

\subsection{Lattice eigenenergies and eigenfunctions}

We numerically diagonalize the full lattice Hamiltonian $\mathcal{H}_{\text{lattice}}$ using parameters in Table\,\ref{tab:SI_qubit-param}:
\begin{equation*}
\begin{split}
 \mathcal{H}_{\text{lattice}}/\hbar & = \sum_{i}{J_{i,i+1} (a_i^\dagger a_{i+1} + a_i a_{i+1}^\dagger)} \\ 
 &+ \sum_{i}{J_{i,i+2} (a_i^\dagger a_{i+2} + a_i a_{i+2}^\dagger)} \\ 
 & +\sum_i{\frac{U_{2}^{q_i}}{2}n_i(n_i-1)} \\
 &+\sum_i{\frac{U_{3}^{q_i}}{6}n_i(n_i-1)(n_i-2)} 
\end{split}
\end{equation*}
The resulting lattice spectrum is used in Fig.\,\ref{fig:fig2}b, Fig.\,\ref{fig:fig4}a, Fig.\,\ref{fig:fig4}b (vertical arrows), while the single-particle wavefunctions are used in Fig.\,\ref{fig:fig3}c.

\subsection{Tunneling spectra}

We use Eqn.\,\ref{eqn:1} and \ref{eqn:2} to calculate the tunneling spectra (solid lines) in Fig.\,\ref{fig:fig2}d and \ref{fig:fig2}h. The excitation spectra (Eqn.\,\ref{eqn:1}) are calculated using the lattice eigenvalues and eigenfunctions above. The frequency Kernel is set to be a Lorentzian with width $\kappa_r$: $K = K(\omega - \omega') = 1/[1+4(\omega - \omega')^2/\kappa_r^2]$. The many-body state $\psi$ is taken to be the initial state: the vacuum state and the $n=1$ Mott insulator, for the quasi-particle and quasi-hole spectroscopy experiment, respectively. We use the experimental parameters: $\kappa_\text{S/D} = \kappa_r = 2\pi\times 1.5$\,MHz, and $g_\text{S/D} = 2\pi\times 0.25$\,MHz. 

This can be viewed as evaluating $\frac{d}{dt} N(\omega, i, t) |_{t\approx0}$, at a time when the probe resolution is no longer Fourier-limited, but when the exponential saturation due to the non-linear spectra has not begun. The result is plotted in Fig.\,\ref{fig:fig2}d and \ref{fig:fig2}h. They are in excellent agreement with the measured spectral weight extracted from the short-time linear slope of exponential fits to $\mathit{\Delta}N(t)$, i.e $\frac{d}{dt}\mathit{\Delta}N |_{t=0}$ (note that $\mathit{\Delta}N$ and $N$ only differ by a constant).

\subsection{Particle number dynamics $\mathit{\Delta}N(t)$}

\begin{figure}[!t]
    \includegraphics[width=0.95\columnwidth]{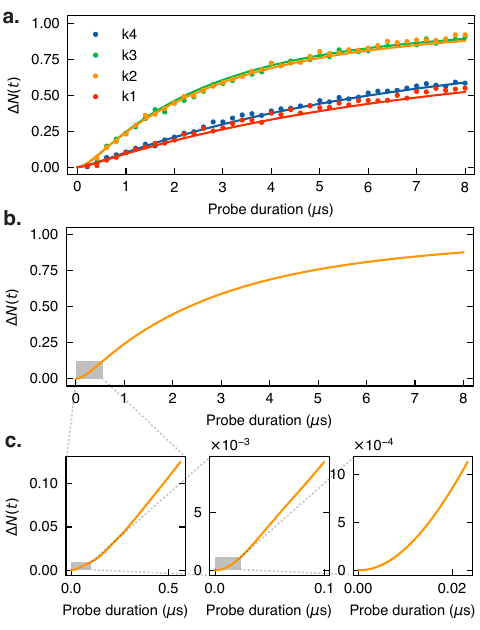}
    \caption{Numerical simulation of the probe-induced population change in the lattice $\mathit{\Delta}N(\delta_\text{D},t)$. Dots in (a) are experimental data in Fig.~\ref{fig:fig2}e; solid lines are numerical results. (b) Numerical results when the probe is resonant with $k_3$, zoomed in to show early time dynamics in (c).
    }
    \label{fig:figSI-time-dyn}
\end{figure}

Here, we show numerical simulations of the particle number as a function of the probe duration to help further support the discussion in Sec.\,III of the main text. As a concrete example, we focus on the scenario in Fig.~\ref{fig:fig2}e, where the initially empty lattice is tunnel coupled to the particle source at site 1. We simulate the dynamics of the system using a Lindblad master equation:
\begin{equation*}
\begin{split}
\dot \rho 
& = -i[\mathcal{H}_\text{coupled}, \rho] + \kappa_r\mathcal{D}[b] \\
& + \sum_i \Gamma_i(1+n^q_\text{th})\mathcal{D}[a_i] + \sum_i \Gamma_i n^q_\text{th}\mathcal{D}[a_i^\dagger]
\end{split}
\end{equation*}
The Hamiltonian includes the lattice, the probe resonator, and the effective coupling between the probe and site 1 induced by the blue-sideband flux modulation,
\begin{equation*}
 \mathcal{H}_{\text{coupled}}/\hbar = \mathcal{H}_{\text{lattice}}/\hbar - \delta_\text{S} b^\dagger b +g_\text{S}(a^\dagger_1 b^\dagger + a_1b),
\end{equation*}
while the collapse operators include the bath resonator decay $\sqrt{\kappa_r}b$ and the intrinsic lattice relaxation towards the equilibrium thermal ground state $\sqrt{\Gamma_i(1+n^q_\text{th})}\,a_i$ and $\sqrt{\Gamma_i n^q_\text{th}}\,a_i^\dagger$, where $\Gamma_i = 1/T^{q_i}_1$. 
Specifically for experiments in Fig.\,\ref{fig:fig2}, the lattice dynamics are restricted to the ground band with at most one particle per site. In addition, the population in the bath resonator remains small in the weak coupling limit. Hence the master equation can be solved by truncation all sites and resonators to only two levels. 

The resulting $\mathit{\Delta}N(t)$ is shown in Fig.\,\ref{fig:figSI-time-dyn}b for a particular resonant probe frequency (mode $k_2$), and shown again in Fig.\,\ref{fig:figSI-time-dyn}a together with the measured data (dots) from Fig.\,\ref{fig:fig2}e. We find excellent agreement between the experiment and the numerical results obtained without any free parameters. In the numerical simulation, we start with the lattice in the ground state with finite thermal population $n^q_\text{th} = 0.05$ on each site and the probe resonator in its ground state; the probe detunings for the four resonances ($k_4$ to $k_1$) are $\delta_\text{S} = 2\pi\times \{-9,-4,3.5,10.5\}$\,MHz.

The numerically simulated dynamics shown here is a good illustration of the discussion in Sec.\,III of the main text on the time dependence of $\mathit{\Delta}N(t)$. When zoomed in to small probe durations (Fig.\,\ref{fig:figSI-time-dyn}c), we see clearly the quadratic scaling of $\mathit{\Delta}N(t)$ with $t$ at early times up to tens of ns. After that, $\mathit{\Delta}N(t)$ transitions into the exponential saturation towards 1 at timescales of a few $\mu$s. As we already estimated, the population change during the early time regime is only at a few percent level, and therefore is not significant in the measured data. This allows us to fit the measured data directly to an exponential function and extract the spectral weight from the slope of the fitted exponential at early times.

As a side note, for very large lattices where the many-body spectrum becomes quasi-continuous, the dynamics of population change when the lattice is coupled to a narrow band probe can be described by Fermi's Golden rule, resulting in a linear scaling $\mathit{\Delta}N(t) \propto t$. See, for example, the discussion in Ref.\,\cite{Kantian2015-sy}.